\documentclass[twocolumn]{article}
\usepackage{relsize}
\usepackage{algorithm}
\usepackage{algpseudocode}
\usepackage[T1]{fontenc}
\usepackage[margin=1in]{geometry}
\usepackage{amsmath}
\usepackage{booktabs}
\usepackage{graphicx}
\usepackage{color}
\usepackage{tikz}
\begin{document}
\newcommand{\z}[1]{{\ttfamily#1}}
\newcommand\concat{\mathbin{|\kern-.2ex|}}

\thispagestyle{empty}
\title{On the Practicality of Dynamic Updates in Fast Searchable Encryption}
\author{Steven Willoughby \\ Portland State University}
\date{Spring 2019}

\twocolumn[
 \begin{@twocolumnfalse}
  \maketitle
  \begin{abstract}
Searchable encrypted (SE) indexing systems are a useful tool for utilizing cloud services 
to store and manage sensitive information. However, much of the work on SE systems to 
date has remained theoretical.
In order to make them of practical use, more work is needed to develop optimal protocols 
and working models for them. 
This includes, in particular, the creation of a working update model in order to maintain 
an encrypted index of a dynamic document set such as an email inbox.
I have created a working, real-world end-to-end SE implementation that satisfies these needs,
including the first empirical performance evaluation of the dynamic SE update operation.
In doing so, I show a viable path to move from the theoretical concepts described by previous
researchers to a future production-worthy implementation and identify issues for follow-on
investigation.
  \bigskip
  \end{abstract}
 \end{@twocolumnfalse}
]

\section{Introduction}
There are many situations and contexts wherein users of information systems need to collect,
store, search, and retrieve large amounts of information. When the collection of data is large
enough or needs to be available to multiple geographically-separated users, an attractive option
may be to host the document repository on a cloud service provided by a third party.

While this allows the users to utilize the service's data centers and network connections to provide
a robust platform to host their data, it opens a number of very serious security and privacy
concerns if the data being hosted are in any way sensitive, since the hosting service may not 
necessarily be trusted to protect that information from their own personnel or others.

Consider, for example, an organization which uses such an externally-hosted searchable repository
to manage confidential pre-release product design documentation, or financial information belonging
to the organization. Worse, consider if the data were to contain personal information about 
employees or customers which would have expensive and disruptive effects on people's lives if it
were to be leaked to unauthorized parties.

The obvious solution is to encrypt the data, so that they may be stored on the untrusted server
in a form that cannot be understood by anyone but authorized personnel.  This solves the problem
of protecting the data at rest on the server. However, since the index must be decrypted in order
to search within it, we must take one of two approaches: either provide decryption keys to the
server in order to decrypt and search server-side, or download the entire index to the client for
the decryption and search to be performed there. The former approach is not desirable because we
have already established that the hosting provider may not be authorized to see the data nor trusted
to protect it from unauthorized access. The latter is less than practical due to the amount of
data which must be copied to users' local systems. These client systems may not have sufficient
storage or processing power\footnote{We must accept in these modern times that client computing
platforms may well include cell phones and low-power notebooks in addition to more traditional
computing platforms.} and the data may well be unreasonably large to transmit repeatedly---it
may be hundreds of megabytes, gigabytes, or terabytes depending on the amount of 
indexed data.

Ideally, we desire to have a method whereby the server can facilitate searches within an index
of interesting keywords from the document repository, then report back with a list of documents 
containing the requested keywords (allowing the user to then retrieve those documents, locally
decrypt them, and make use of their contents), all without the server having the ability to actually
read the document index itself (since that provides a great deal of insight into the contents
of each indexed document). In fact, the server should not even be able to understand
what keywords it was searching for (since that provides insight into the nature of the documents
and what the users are looking for), or what documents were in the result list of each search.

While that may seem to be an impossible expectation, in reality we can find
an acceptable middle ground which allows efficient server-side searching without divulging any
direct information about the details of the search terms or results. The price paid for this,
however, is that a determined hostile observer (perhaps the hosting provider themselves) could
analyze patterns of input and output over time which will ``leak'' useful information from which
some amount of the protected data may be inferred.

Building on the foundational work of previous researchers in this field, I have created a dynamic
update capability which allows an SE index to accumulate new documents over time, whereas previous
implementations were primarily focused on a one-time generation of an SE index for a static document
set. I also moved beyond the previous theoretical treatments of this subject by adding empirical performance
evaluation of my new update mechanism using a typical \textsc{tcp/ip} client-server architecture. Based
on this work I identified some considerations for future optimization work.

\section{Definitions and Nomenclature}
In this paper I will use the terminology set out by Curtmola, et al.\@ \cite{curtmola} which is also used by other authors, notably Demertzis and Papamanthou, \cite{demertzis}
for the sake of consistency with established work on this topic.
Basic notation and symbology is summarized in Table~\ref{tbl:notation}.
\begin{table*}
 \begin{center}
  \begin{tabular}{ll}\toprule
   \bfseries Notation & \bfseries Meaning \\\midrule
   $a\concat b$			& Concatenation of strings $a$ and $b$		\\
   $\left|X\right|$		& Cardinality of set $X$			\\
   $x\oplus y$			& Bitwise exclusive-or of $x$ and $y$		\\
   $x\xleftarrow{\$}X$  	& Element $x$ sampled uniformly from set $X$ 	\\
   $x\leftarrow\cal A$ or ${\cal A}\rightarrow x$		
				& Output $x$ from algorithm or function $\cal A$\\
   $\Delta=\{w_1,w_2,\ldots,w_d\}$ 	& Dictionary of $d$ words in an index	\\
   ${\cal D}=\{D_1,D_2,\ldots,D_n\}$
				& Set of $n$ documents whose words are indexed \\
   ${\cal D}(w)$		& List of all documents containing word $w$ 	\\
   $A_i[x]$                     & Bucket $x$ of level $i$ in index storage array $A$ \\
   $\lambda$                    & Bit length of encryption keys \\
   $L$        			& Locality of the index \\
   ${\cal L}=\{i_1,i_2,\ldots,i_s\}$& Set of $s$ storage levels in use for the index \\
   $N$				& Number of stored $(w,\textsf{id}(D))$ tuples in index \\
   $o$				& Order of a SE index, related to its storage capacity \\
   $s$                          & Number of actually stored index levels \\
   $c\leftarrow\textsf{Enc}(K,m)$	
				& Encryption function with key $K$ and plaintext message $m$ \\
   $m\leftarrow\textsf{Dec}(K,c)$	
				& Decryption function with key $K$ and ciphertext message $c$ \\
   $y\leftarrow\textsf{F}(K,x)$ & Pseudo-random function with key $K$ and data $x$ \\
   $x'\leftarrow\textsf{H}(x)$	& Collision-resistant hash function taking data $x$ \\
   $\textsf{id}(D)$		& Unique identifier for document $D$		\\
   $\varepsilon$		& Empty string or unused storage location \\
   \z{000C}                     & Hexadecimal values are shown in fixed-width type \\
   \bottomrule
  \end{tabular}
  \caption{Summary of Notation Used in This Paper\label{tbl:notation}}
 \end{center}
\end{table*}

Central to this topic is the notion of a collection of documents for which the user wishes to 
maintain a searchable encrypted index. Following Curtmola, et al.'s nomenclature,
let $\Delta$ be a dictionary of all ``interesting'' words
in all documents, i.e., $\Delta=\{w_1,\ldots,w_d\}$ where $d$ is the number of unique 
interesting words. If $2^{\Delta}$ is the power set of all possible documents containing 
words $w\in\Delta$, then we will consider a set ${\cal D}\subseteq2^{\Delta}$ which is the 
specific set of $n$ documents being indexed in some particular instance of searchable encrypted 
index being discussed.

Each such document has a unique identifier by which it can be fetched from its storage location.
Let $\textsf{id}(D)$ be the identifier for some arbitrary document $D$. 
Further, let ${\cal D}(w)=\{\textsf{id}(D)\ \forall D\in{\cal D} \mid w\in D\}$ be the set of unique identifiers for all documents in our indexed collection which contain the word $w$.
(Curtmola, et al.\@ use the notation $\textbf{D}$ and $\textbf{D}(w)$ instead of
$\cal D$ and ${\cal D}(w)$ respectively).

For my work which builds primarily on the work by Demertzis and Papamanthou, \cite{demertzis} I will also use the following 
nomenclature from their work: 
Let $\lambda$ be the security parameter (in practical terms, the encryption key length in bits),
such that each key $k_i$ is generated using a cryptographically secure random number source, i.e., 
$k_i\xleftarrow{\$}\{0,1\}^{\lambda}$. 

Also let $N$ be the number of entries stored in the SE, 
where \emph{entry} is a word which here means a unique tuple $(w,\textsf{id}(D))$ mapping 
an indexed keyword $w$ to the identifier of a document $D$ containing that word. 
Thus, we have \[N=\sum_{\forall w\in\Delta}\left|{\cal D}(w)\right|.\]

As we shall see, Demertzis and Papamanthou \cite{demertzis} posit a storage array arranged in
tiered \emph{levels} of varying sized storage \emph{buckets}.

Let $\ell=\lceil\log_2 N\rceil$ be the number of levels of index storage which would be employed in
this model.

Let $s\le\ell$ be a configurable number of tiers which will actually be stored on the server
(to save space since not all indexes will have values actually assigned to all possible levels),
and $\cal L$ be the set of $s$ storage levels allocated for the SE.

This SE model supports the notion of \emph{locality} where data associated with the same
keyword are placed in 1 or more co-located areas in the data store.
Let $L$ be the user-configurable locality such that specifying $L>1$ allows each
indexed term to be stored in multiple non-contiguous storage areas, facilitating parallelization
of search operations within the index.
These levels of storage are implemented in storage arrays $A_i$ where $i\in\cal L$. Each
level is further partitioned into \emph{buckets}. Bucket $x$ of array $A_i$ is denoted $A_i[x]$.

I will refer to a few standard functions, as follows. 
Let $\textsf{F}$ be a pseudo-random function (\textsc{prf})
$\textsf{F}:\{0,1\}^*\times\{0,1\}^*\rightarrow\{0,1\}^*$, which emits a deterministic pattern
of bits based on the values of its two inputs (key and data), but whose output is indistinguishable
from random bits if those inputs are not known.
Let \textsf{Enc} and \textsf{Dec} be
\textsc{cpa}-secure\footnote{A \textsc{cpa}-secure cipher is one which is resists chosen-plaintext
attacks.}
symmetric encryption functions 
$\textsf{Enc}:\{0,1\}^*\times\{0,1\}^{\lambda}\rightarrow\{0,1\}^*$
and
$\textsf{Dec}:\{0,1\}^*\times\{0,1\}^{\lambda}\rightarrow\{0,1\}^*$
(such that $\textsf{Dec}=\textsf{Enc}^{-1}$)
which take $\lambda$-bit keys to transform arbitrary-length bit strings to another arbitrary-length ciphertext and back again.
Finally, let $\textsf{H}:\{0,1\}^*\rightarrow\{0,1\}^b$ be a cryptographically strong one-way 
hash function which outputs $b$ bits of digest from its input data of arbitrary length. This
function must be collision resistant.

To the above notation I add the concept of the \emph{order} of an index, 
which gives us a useful way to organize a collection of various-size SE indexes. 
For this research, I chose to assume the order $o$ of an index to be 
$o=\ell=\lceil\log_2 N\rceil$ with the intention that it would yield a reasonable pattern 
of varying sizes of indexes to avoid expensive large-index merge operations as long as 
reasonably possible.

\section{Basic Principles of SE}
Here, and throughout the rest of this paper, the term \emph{client} shall refer to the
system a user of the SE system employs to initiate searches or add new documents to the
SE index. It is a trusted system under the control of an authorized user. Encryption keys
may be employed on it, and plain-text search terms and results may be known to it.

The term \emph{server} shall refer to the remote system on which the encrypted documents
and the SE indexes are stored. This system is not allowed to see any decryption keys nor
to see the plaintext search terms nor results.

The essential principle on which SE is based is that, given an index $\cal I$ mapping 
a set $\Delta$ of interesting keywords from a document repository $\cal D$, we must represent
$\cal I$ in some opaque fashion such that it can be stored on an untrusted server without
anyone being able to glean information about $\cal D$ by examining $\cal I$, even given an
arbitrarily large amount of time to analyze $\cal I$. This implies the use of a one-way
cryptographically strong hash function, since that will provide a way to derive an opaque value
to represent a value in the index without a reliable way to reverse the encoding function to 
obtain the original value again.

If we can then use the same hash function to encode the client-side search terms we can
match them on the server to the encoded entries in $\cal I$ without revealing the original
search terms directly.

To illustrate this concept, consider a document repository which contains five documents,
specifically, the first five volumes of Douglas Adams' magnum opus \emph{The Hitchhiker's Guide
to the Galaxy}. 
These volumes are separately encrypted and stored on the server. Each is
assigned a document ID as shown in Table~\ref{tbl:hhggdocs}.
\begin{table*}
 \begin{center}
  \begin{tabular}{rl}\toprule
   \bfseries ID & \bfseries Document Title\\\midrule
    3 & \emph{The Hitchhiker's Guide to the Galaxy} \\
    5 & \emph{The Restaurant at the End of the Universe} \\
    8 & \emph{Life, the Universe, and Everything} \\
   12 & \emph{So Long, and Thanks for All the Fish} \\
   15 & \emph{Mostly Harmless}\\\bottomrule
  \end{tabular}
  \caption{Example Document Repository $\cal D$\label{tbl:hhggdocs}}
 \end{center}
\end{table*}

We identify a set $\Delta$ of all words in these documents we find interesting for our purposes.
Say, for example, $\Delta=\{$
\z{Arthur},
\z{dolphin},
\z{Fen\-church},
\z{hooloovoo},
\z{krikkit},
\z{Zaphod}
$\}$. (Obviously, in a full production repository the list of interesting words would be
orders of magnitude greater than this trivial example.)
If we make a list of all documents in $\cal D$ in which each of the words in $\Delta$ appear,
we find the following associations of each keyword $w\in\Delta$ to a set of document IDs
$D(w)$:
\begin{align*}
 \text{\z{Arthur}} 	&\rightarrow\{3, 5, 8, 12, 15\}\\
 \text{\z{dolphin}} 	&\rightarrow\{3,       12    \}\\
 \text{\z{Fenchurch}} 	&\rightarrow\{         12, 15\}\\
 \text{\z{hooloovoo}} 	&\rightarrow\{3              \}\\
 \text{\z{krikkit}} 	&\rightarrow\{      8, 12    \}\\
 \text{\z{Zaphod}} 	&\rightarrow\{3, 5, 8, 12, 15\}\\
\end{align*}

From these associations we generate an index $\cal I$ which is a collection of tuples 
$(w,\textsf{id}(D))$. Specifically, we get:\label{tuples}
$(\text{\z{Arthur}}, 3)$,
$(\text{\z{Arthur}}, 5)$,
$(\text{\z{Arthur}}, 8)$,
$(\text{\z{Arthur}}, 12)$,
$(\text{\z{Arthur}}, 15)$,
$(\text{\z{dolphin}}, 3)$,
$(\text{\z{dolphin}}, 12)$,
$(\text{\z{Fenchurch}}, 12)$,
$(\text{\z{Fenchurch}}, 15)$,
$(\text{\z{hooloovoo}}, 3)$,
$(\text{\z{krikkit}}, 8)$,
$(\text{\z{krikkit}}, 12)$,
$(\text{\z{Zaphod}}, 3)$,
$(\text{\z{Zaphod}}, 5)$,
$(\text{\z{Zaphod}}, 8)$,
$(\text{\z{Zaphod}}, 12)$,
and
$(\text{\z{Zaphod}}, 15)$.

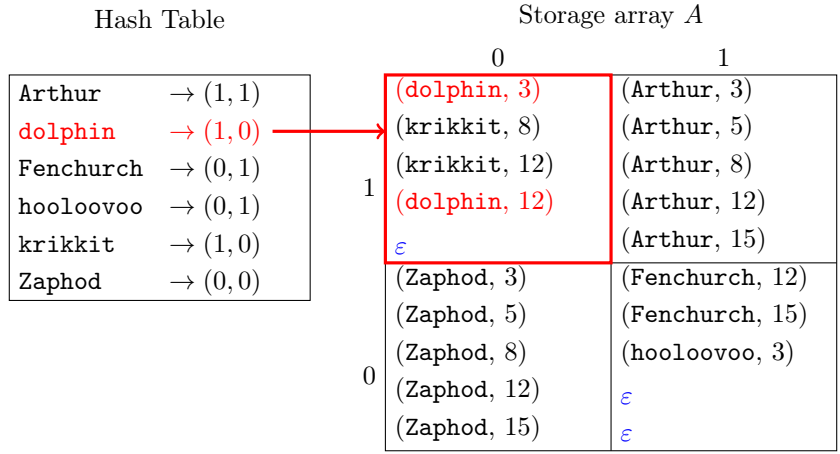
\begin{figure*}
 \begin{center}
  \begin{tikzpicture}
   \draw (0,0) -- (0,5) -- (6,5) -- (6,0) -- cycle;
   \draw (0,2.5) -- (6,2.5);
   \draw (3,0) -- (3,5);
   \node [above] at (3,5.5) {Storage array $A$};
   \node [above] at (-3,5.5) {Hash Table};
   \node [above] at (1.5,5) {0};
   \node [above] at (4.5,5) {1};
   \node [left] at (0,3.5) {1};
   \node [left] at (0,1) {0};
   \draw (-5,2) -- (-1,2) -- (-1,5) -- (-5,5) -- cycle;
   \node [above right] at (0,4.5) {\color{red}(\z{dolphin}, 3)};
   \node [above right] at (0,4.0) {(\z{krikkit}, 8)};
   \node [above right] at (0,3.5) {(\z{krikkit}, 12)};
   \node [above right] at (0,3.0) {\color{red}(\z{dolphin}, 12)};
   \node [above right] at (0,2.5) {\color{blue}$\varepsilon$};
   \node [above right] at (0,2.0) {(\z{Zaphod}, 3)};
   \node [above right] at (0,1.5) {(\z{Zaphod}, 5)};
   \node [above right] at (0,1.0) {(\z{Zaphod}, 8)};
   \node [above right] at (0,0.5) {(\z{Zaphod}, 12)};
   \node [above right] at (0,0.0) {(\z{Zaphod}, 15)};
   \node [above right] at (3,4.5) {(\z{Arthur}, 3)};
   \node [above right] at (3,4.0) {(\z{Arthur}, 5)};
   \node [above right] at (3,3.5) {(\z{Arthur}, 8)};
   \node [above right] at (3,3.0) {(\z{Arthur}, 12)};
   \node [above right] at (3,2.5) {(\z{Arthur}, 15)};
   \node [above right] at (3,2.0) {(\z{Fenchurch}, 12)};
   \node [above right] at (3,1.5) {(\z{Fenchurch}, 15)};
   \node [above right] at (3,1.0) {(\z{hooloovoo}, 3)};
   \node [above right] at (3,0.5) {\color{blue}$\varepsilon$};
   \node [above right] at (3,0.0) {\color{blue}$\varepsilon$};
   \node [above right] at (-5,4.4) {\strut\z{Arthur}};
   \node [above right] at (-5,3.9) {\color{red}\strut\z{dolphin}};
   \node [above right] at (-5,3.4) {\strut\z{Fenchurch}};
   \node [above right] at (-5,2.9) {\strut\z{hooloovoo}};
   \node [above right] at (-5,2.4) {\strut\z{krikkit}};
   \node [above right] at (-5,1.9) {\strut\z{Zaphod}};
   \node [above right] at (-3,4.4) {\strut$\rightarrow(1,1)$};
   \node [above right] at (-3,3.9) {\color{red}\strut$\rightarrow(1,0)$};
   \node [above right] at (-3,3.4) {\strut$\rightarrow(0,1)$};
   \node [above right] at (-3,2.9) {\strut$\rightarrow(0,1)$};
   \node [above right] at (-3,2.4) {\strut$\rightarrow(1,0)$};
   \node [above right] at (-3,1.9) {\strut$\rightarrow(0,0)$};
   \draw [very thick,red] (0,2.5) -- (3,2.5) -- (3,5) -- (0,5) -- cycle;
   \draw [very thick,red,->] (-1.5,4.25) -- (0,4.25);
  \end{tikzpicture}
  \caption{Example Index $\cal I$ Storage (unencrypted)\label{fig:sampleI}}
 \end{center}
\end{figure*}

We store $\cal I$ on disk in two parts: a storage array which holds the actual tuples,
and a hash table which associates each search term $w$ with the location in storage
holding its set of tuples. Setting aside for the moment the finer points of storage
optimization so that we may focus just on the encryption aspect, let us visualize the
storage arrangement of our index $\cal I$ as shown in Figure~\ref{fig:sampleI}.

With such a storage arrangement, if the client wishes to search for keyword $w=\text{\z{dolphin}}$,
the server looks that up in the hash table, finding that the tuples to satisfy the search are
contained in storage array level 1, bucket 0 (which we will designate $A_1[0]$). Looking in that
bucket, we find (among other things that happen to be stored there as well) the tuples
$(\text{\z{dolphin}},3)$
and
$(\text{\z{dolphin}},12)$.
From this the server reports the result set $\{3,12\}$ as the set of document IDs where the
word ``\z{dolphin}'' is found. 

\subsection{Encrypting the Index}
To the above trivial storage arrange we now need to add a layer of encryption to obscure the
meaning of the information in $\cal I$ beyond the ability of the server to understand, but in
such a way that the client can use it to get the same search results.

For this encryption, we generate a secret key known only to authorized clients. 
This key $K=(k_1,k_2,k_3)$ has three parts, each of which is created from $\lambda$ random
bits (i.e., $k_i\xleftarrow{\$}\{0,1\}^{\lambda}$).

\begin{figure*}
 \begin{center}
  \begin{tikzpicture}
   \draw (0,0) -- (0,5) -- (6,5) -- (6,0) -- cycle;
   \draw (0,2.5) -- (6,2.5);
   \draw (3,0) -- (3,5);
   \node [above] at (3,5.5) {Storage array $A$};
   \node [above] at (-3,5.5) {Hash Table};
   \node [above] at (1.5,5) {0};
   \node [above] at (4.5,5) {1};
   \node [left] at (0,3.5) {1};
   \node [left] at (0,1) {0};
   \draw (-5,2) -- (-1,2) -- (-1,5) -- (-5,5) -- cycle;
   \node [above right] at (0,4.5) {\color{red}\z{A462910E}};
   \node [above right] at (0,4.0) {\z{3002B257}};
   \node [above right] at (0,3.5) {\z{6B9CB117}};
   \node [above right] at (0,3.0) {\color{red}\z{07B422A7}};
   \node [above right] at (0,2.5) {\color{blue}\z{BA84D75F}};
   \node [above right] at (0,2.0) {\z{8E095BDB}};
   \node [above right] at (0,1.5) {\z{130651E7}};
   \node [above right] at (0,1.0) {\z{78B1C20B}};
   \node [above right] at (0,0.5) {\z{DBB21619}};
   \node [above right] at (0,0.0) {\z{D743999B}};
   \node [above right] at (3,4.5) {\z{DCF582AE}};
   \node [above right] at (3,4.0) {\z{E704FD8D}};
   \node [above right] at (3,3.5) {\z{DB90A5D4}};
   \node [above right] at (3,3.0) {\z{A9633ECF}};
   \node [above right] at (3,2.5) {\z{F25870D9}};
   \node [above right] at (3,2.0) {\z{923F4350}};
   \node [above right] at (3,1.5) {\z{D8D756C2}};
   \node [above right] at (3,1.0) {\z{6F89A58C}};
   \node [above right] at (3,0.5) {\color{blue}\z{872784E9}};
   \node [above right] at (3,0.0) {\color{blue}\z{83A3F977}};
   \node [above right] at (-5,4.5) {\z{183BFF00}};
   \node [above right] at (-5,4.0) {\color{red}\z{38A9039C}};
   \node [above right] at (-5,3.5) {\z{9FB946BA}};
   \node [above right] at (-5,3.0) {\z{296A7C1E}};
   \node [above right] at (-5,2.5) {\z{9F66B745}};
   \node [above right] at (-5,2.0) {\z{89C122B2}};
   \node [above right] at (-3.2,4.5) {$\rightarrow$ \z{3C5C0D95}};
   \node [above right] at (-3.2,4.0) {$\rightarrow$ \color{red}\z{6BF86758}};
   \node [above right] at (-3.2,3.5) {$\rightarrow$ \z{C20E642D}};
   \node [above right] at (-3.2,3.0) {$\rightarrow$ \z{A305B1C5}};
   \node [above right] at (-3.2,2.5) {$\rightarrow$ \z{A54A58BF}};
   \node [above right] at (-3.2,2.0) {$\rightarrow$ \z{0ADE7001}};
   \draw [very thick,red] (0,2.5) -- (3,2.5) -- (3,5) -- (0,5) -- cycle;
   \draw [very thick,red,->] (-1.1,4.25) -- (0,4.25);
  \end{tikzpicture}
  \caption{Example Index $\cal I$ Storage (encrypted)\label{fig:encryptedI}}
 \end{center}
\end{figure*}
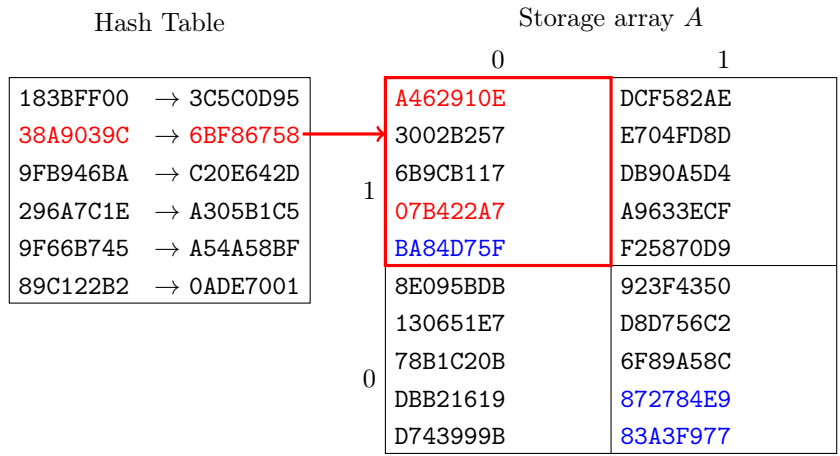

First, given a cryptographically strong one-way hash function \textsf{H},
pseudo-random function \textsf{F}, and encryption function \textsf{Enc} as described above,
we encode the tuples stored in the array $A$ by encrypting the value
$\textsf{id}(D)\concat0^{\lambda}$ using the encryption key
$\textsf{F}(k_3,w)$. In our example, assuming for simplicity that document IDs
are 16 bits and $\lambda=16$, the tuple $(\text{\z{dolphin}},3)$ is encoded by
calculating $\textsf{Enc}(\textsf{F}(k_3,\text{\z{dolphin}}), \text{\z{00030000}})$.
Likewise, the tuple
$(\text{\z{dolphin}},12)$ is encoded by
calculating $\textsf{Enc}(\textsf{F}(k_3,\text{\z{dolphin}}), \text{\z{000C0000}})$.
Assuming these two calculations produce the hex values \z{A462910E} and \z{07B422A7},
and that we carried out corresponding encodings with the other tuples, we would now have
the encrypted storage array shown in Figure~\ref{fig:encryptedI}. Note that we also
filled the empty storage locations with random bits to further obfuscate the index.

It is important to note that each tuple is encrypted with a key that is based on
the search term to which it belongs, so the data there is only recoverable if
one is in possession of that secret key $k_3$ and the search term $w$.

Now that the tuples are encoded, we must encrypt the hash table's keys and values in
a similar fashion. The keys (the search terms) are simply replaced with the results
of hashing them with another secret key: $\textsf{H}(\textsf{F}(k_1,w))$. Thus, search
term ``\z{dolphin}'' would be replaced by $\textsf{H}(\textsf{F}(k_1,\text{\z{dolphin}}))$,
say \z{38A9039C}.

The value associated with ``\z{dolphin}'', is the tuple $(1,0)$ which means that
the entries for that keyword are to be found in $A_1[0]$ (storage level 1, bucket 0).
We represent location $A_i[x]$ as a single numeric value $i\concat x$
(in this case the hex value \z{00010000}). This is encoded in the hash table as
$[i\concat x]\oplus\textsf{H}(\textsf{F}(k_2,w))$.
Again, note that the search term $w$ and a secret key are part of this encryption scheme.
Supposing this gives the result \z{6BF86758}, and continuing this for the rest of the table,
we get the completely encrypted index shown in Figure~\ref{fig:encryptedI}. The values 
where the entries for our example term ``\z{dolphin}'' are encoded in $\cal I$ are highlighted
in red.

Now if we wish to search for a word like ``\z{dolphin}'', we generate a \emph{search token}\label{token}
$T=(t_1,t_2,t_3)$ by providing the portion of the encoding operations requiring knowledge
of the secret values, sending to the server only the output from \textsf{F} which it can use
to complete the hashing and decryption without divulging the actual keys or search terms:
$t_1=\textsf{F}(k_1,w)$,
$t_2=\textsf{F}(k_2,w)$,
and
$t_3=\textsf{F}(k_3,w)$.

The server, upon receiving the client's search token $T$, calculates 
$\textsf{H}(t_1)$ and gets \z{38A9039C}. Looking at the hash table in Figure~\ref{fig:encryptedI}
we see that this is a key stored there, associated with value \z{6BF86758}.
The server then calculates $\text{\z{6BF86758}}\oplus\textsf{H}(t_2)$ to get the result \z{00010000}. Although the server never knew the search term $w$, it was given just enough information in $T$
to determine that the answer to that query is to be found in storage location $A_1[0]$. $T$ does
not provide any information to decode any other hash table entries since they were encoded using
different values of $w$.

Now the server knows that some of the values stored in $A_1[0]$ can be decrypted using 
the key $\textsf{H}(t_3)$. Running the contents of $A_1[0]$ through this decryption, it gets
the results \z{00030000}, \z{1AED5898}, \z{EF00F293}, \z{000C0000}, and \z{923BF508}.
Since any valid entry has $0^{\lambda}$ bits appended, the server knows that only the first 
and fourth values were correctly decrypted by the key it was given, so the result reported
to the client is the set of document IDs $\{3,12\}$.

Note that when we set locality $L>1$, we must allow for multiple buckets to hold the tuple lists
for any given keyword, so the actual calculations for the hash table keys and values includes a 
counter $c\in\{0,1,\ldots,L-1\}$. The key is actually encoded as 
$\textsf{H}(\textsf{F}(k_1,w)\concat c)$ and the value as
$[i\concat x]\oplus\textsf{H}(\textsf{F}(k_2,w)\concat c)$.

\section{Prior Work}
In their seminal work on the subject, Song, Wagner, and Perrig \cite{song} laid out the essential idea for SE indexing
for the first time. From this beginning almost two decades ago, other researchers have further developed and extended
these initial concepts in order to improve functionality, security, and performance.

One approach explored by Pinkas and Reinman \cite{pinkas} as an alternative to SE was to leverage the concept of 
oblivious \textsc{ram} (\textsc{oram})---a
specially-arranged memory system originally proposed by Goldreich and Ostrovsky \cite{goldreich} which 
has the property that ``the sequence of memory accesses \dots\ reveals no information about the input \dots, beyond
the running-time for the input.'' Pinkas and Reinman sought to use this aspect of \textsc{oram} to hide the nature of
the calculations used to search through an encrypted index to thwart attempts at cryptanalysis or other means of obtaining
confidential details of the index. Unfortunately, this approach is very expensive compared to more practical software-only
solutions described here.

As these software SE systems were developed, they were primarily implemented as in-memory schemes. Cash, et al.\@ 
\cite{cash} note that this approach did not scale effectively as the repository size expanded into the near-terabyte 
range and beyond. As index entries may be scattered throughout the index, the amount of data transmitted back to the 
user for local decryption multiplies with the database size. Cash and his co-authors proposed refinements which resulted 
in greater locality of the encrypted index entries, guaranteeing that entries matching a given search term cluster near 
each other in the index, thus reducing the number of encrypted index blocks which must be sent.

Cash and Tessaro \cite{cash2} continued improving their previous SE schemes, working on maximizing data locality---the number of non-contiguous storage groups from which the server reads data to satisfy a given search query. They note---and go
on to formally prove---how this optimization runs counter to the need to reduce the size of the index storage on the server.

Building further on that research, Asharov, et al.\@ \cite{asharov} created SE schemes with improved read efficiency 
(they report $O(\log n)$ and $O(\log\log n)$) and demonstrated that it will always be the case that to achieve 
either maximal read efficiency and locality it will be necessary to sacrifice one to achieve the other.

Finally, Demertzis and Papamanthou \cite{demertzis} improved on these earlier efforts by developing a scheme which
provides reasonable locality, including controls the repository owner may adjust to ``tune'' the storage to be 
maximally efficient for the type of data being indexed. 

My research is directly based on the work of Demertzis and Papamanthou, whose scheme I extended to include multiple
index collections and dynamic updates.

\subsection{Security of SE Systems}
The observation above that SE systems will ``leak'' information over time from which an observer can infer confidential
information raises the question of how much information leakage is acceptable.
This issue has been explored at length by previous researchers. Song, Demertzis, and their colleagues who developed
their respective SE implementation models (e.g., \cite{song,demertzis}) provided with them formal proofs of the security
of their encryption schemes. This was important to establish the trustworthiness of the SE concept in general.

Following on from this foundation, Naveed, Kamara, and Wright \cite{naveed} along with Zhang, Katz, and Papamanthou
\cite{zhang} studied various attack scenarios and found that
it was possible for a determined observer to eventually decrypt a significant amount of information out of an
encrypted database stored on an untrusted server. These findings helped drive Demertzis and Papamanthou to
develop more cryptographically robust encryption schemes which I also used for my work, and prompted me to 
seek a means of periodically invalidating accumulated inferences an observer may have gleaned as part 
of my update procedure. 
 
\subsection{Locality Optimization}
As noted above, early SE research posited in-memory solutions for the sake of theoretical exploration, but this 
presented a roadblock to adapting SE systems for real-world applications as it didn't allow the indexes to scale
up to the data sizes needed in the real world.
To address this, a number of storage strategies were proposed by Cash, et al., \cite{cash,cash2} but 
these often ran into difficulties. 
For example, the practice of obfuscating the layout of the index by permuting index data throughout 
the storage area came at the expense of having related data clustered to make reads more efficient.

Demertzis and Papamanthou \cite{demertzis} proposed one improvement which I found advantageous enough to
base my own work upon. 
Given some array $A$ of storage locations on the server, this is organized into tiered \emph{levels}
$A_0, A_1, \ldots, A_\ell$,
where each level $A_i$ consists of a number of \emph{buckets} $A_i[0], A_i[1], \ldots, A_i[q_i]$ in which
we will store document IDs.

At each level, the bucket sizes increase exponentially. For example, one level would hold buckets 
containing 2 references, the next level would hold buckets of size 4, the next 8, the next 16, and so forth. 
The documents themselves are stored as encoded $(w,\textsf{id}(D))$ tuples as described above on p.~\pageref{tuples}.

This arrangement nicely facilitates our need to populate the index with document IDs where the number of IDs
matching any given keyword varies widely from the number matching another keyword, while allowing us to 
co-locate these tuples within $L$ buckets for efficiency. By adjusting the value of $L$ at index creation time,
the SE administrator can reduce the locality of the tuple storage but gain the ability to split up searches
into parallel tasks.

They also introduced the optimization parameter $s$ which allows an index to be built with only a subset of
levels actually used. Specifically, for $s=\ell$, all levels are utilized, with each level $i$ containing
buckets sized to hold $2^i$ tuples. If $s$ is reduced to some value $1\le s\le\ell$, however, the set of 
actual levels utilized will be \[{\cal L}=\{\ell, \ell-p, \ldots, \ell-(s-1)p\}\] and the tuples will be 
stored in the nearest actual level to the one it would have been assigned if all were allocated.

\section{My Contributions}
I focused my work in two specific areas: to create a working production-scale SE implementation based on
Demertzis and Papamanthou's model, \cite{demertzis} and then to develop a system to add more information
to the SE index over time. Their procedure for building a new SE index is summarized in Algorithm~\ref{alg:setup}.

\begin{algorithm*}
 \begin{center}
  \begin{algorithmic}
   \Procedure{Setup}{$k,\cal D$}\Comment{$L,s$ (locality and stored levels) are publicly known parameters}
   \State Let $\Delta$ be the list of all ``interesting'' keywords in document set $\cal D$.
   \State Let $N=\sum_{\forall w\in\Delta}|{\cal D}(w)|$; $\ell=\lceil\log N\rceil$; and $p=\lceil\ell/s\rceil$.
   \State Let ${\cal L}=\{\ell,\ell-p,\ldots,\ell-(s-1)p\}$. 
   \If{$L>1$} 
    \State ${\cal L}\gets{\cal L}\cup\{0\}$
   \EndIf
   \State $\forall i\in\cal L$ organize storage level $A_i$, divided into buckets $A_i[x]$.
   \For{each keyword $w\in \Delta$ in random order}
   	\State Find adjacent $i,j\in{\cal L}:L2^j<|{\cal D}(w)|\le L2^i$.
   	\State Split ${\cal D}(w)$ into a set of chunks $C_w$. Set $c=0$.
   	\For{each chunk $v\in C_w$}
   		\State $c\gets c+1$.
   		\State Let $A$ be buckets in $A_i$ able to hold chunk $v$.
   		\State Pick one bucket $A_i[x]$ from $A$ randomly; store $v$ in it.
   		\State Add $H(F(k_1,w)\concat c)\Rightarrow [i\concat x]\oplus H(F(k_2,w)\concat c)$ to hash table.
   	\EndFor
   \EndFor
   \State Permute and encrypt entries in ${\cal L}$; fill HT with random data.
   \EndProcedure
  \end{algorithmic}
  \caption{Build an Index (summarized from Demertzis and Papamanthou \cite{demertzis})}\label{alg:setup}
 \end{center}
\end{algorithm*}

\subsection{Real-World Implementation}
I investigated two avenues for implementing a remotely hosted SE indexing system. 
The first was to implement an indexing scheme that maintained its indexes in local files. This was done with
some straightforward Python programs:
\begin{itemize}
 \item	\z{genkeys} generates a set of cryptographic keys $K=(k_1,k_2,k_3)$ for use by the client to
	encrypt and decrypt index information.
 \item	\z{buildindex} reads a collection of documents, extracting a list of words from each.
	These words are then encoded into an encrypted index stored as a simple \textsc{dbm} database.
 \item	\z{search} takes a search query from the user, looks that up in the SE index, and reports
	the matching list of document IDs to the user.
\end{itemize}
Since these operate on local files, all scripts are considered ``client-side'' in terms of having access
to secret keys. In this model, I had in mind an implementation where another operating system layer---transparent
to the SE code---handles remote hosting of the files. My choice for this was the InterPlanetary File System
(\textsc{ipfs}), \cite{ipfs} which provides a distributed filesystem between clients, so each user
sees a copy of the same underlying files, allowing a purely localized operation. 

However, while that provides for simplicity of SE implementation, it comes at too high a cost for the
widest audience since it requires substantial local data storage to be available on every client.
It did, however, serve to demonstrate the correctness of the basic SE operations themselves before adding
the extra complexity of network operations.

From there I switched to a traditional client-server model, defining a hard separation of duties between the data host
(which may be remote and untrusted) and the local client. This is implemented in a new set of Python programs:
\begin{itemize}
 \item	\z{fseserver} runs on the server system to carry out requests on behalf of clients. This manages
	the \textsc{dbm} databases which comprise the SE index.
 \item	\z{buildindex\_client} works as \z{buildindex} does but rather than building local database files,
	it encrypts a new index and sends it to the server for remote storage.
 \item	\z{search\_client} takes a search query from the user, computes a search token $T$ as described 
	on p.~\pageref{token}, and passes that to the server. It relays the search results from the server
	back to the local user.
\end{itemize}

When designing the client-sever protocol for my implementation, one of my significant design goals was to
allow large blocks of raw binary data since so much of the index is encrypted and needs to be sent as whole buckets
at a time. This way I would not waste resources encoding and decoding the binary blocks as, e.g., base-64.

\subsection{Dynamic SE}
The key intention of my work on SE systems was to find a practical implementation of \emph{dynamic} SE indexing.
In today's world it seems quite likely that an SE indexing system would be gainfully employed to help users search
through a fluid stream of communication, such as an email inbox or the conversation history of an official
chat service such as those used for customer support by some companies.

Most of the work to this point has referenced the idea of building an index $\cal I$ from a set of documents
$\cal D$ and set of search terms $\Delta$ in a single indexing operation. Once $\cal I$ is built, it is then
searched any number of times but there is no notion of $\cal I$ changing over time. Indeed, the way $\cal I$ is
constructed in the first place depends on values such as $N$ (the number of $(w,\textsf{id}(D))$ tuples stored)
and the distribution of words $w\in\Delta$ throughout the data set and the number of documents ${\cal D}(w)$ for
each. If those values change, the internal arrangement of the whole index may be different.

This implies that updating $\cal I$ over time is necessarily a matter of rebuilding it again from scratch. However,
this is obviously untenable for large indexes (e.g., when adding a single email message to an existing index holding
10 million messages already).

Cash, et al.\ \cite{cash} discuss their own approach to dynamic SE schemes and the limitations they and others
encountered. They note that prior schemes ``had an impractically large index or leaked [information allowing
the server to learn] the pattern of which keywords appear in which documents\dots\ which is a severe form of
leakage.'' They go on to improve on that but make the assumption that full index rebuilds will be performed
periodically and that deletions from the index will be rare.  However, the approach they took does not lend itself
to the tiered architecture I am working with. This prompted me to implement a new dynamic update system which
is compatible with the tiered organization so I can retain the optimizations afforded by that structure.

Demertzis and Papamanthou \cite{demertzis} do discuss dynamic updates to SE systems, but only to a limited
extent. 
While acknowledging the shortcomings of previous attempts, their proposal was sketched out in basic terms: 
``The main idea is that we organize $n$ sequential updates to a collection of \dots\ independent encrypted 
indexes\dots.  [For each $(w,{\cal D}(w))$ tuple mapping a search word to a document ID,] 
the data owner initializes a new SE scheme by creating a new SE index that contains only the specific
tuple, [that] is subsequently uploaded to the untrusted server. Whenever two indexes of the same size
$t$ are detected there [sic] are downloaded by the data owner, decrypted and merged to form a sew SE index of size $2t$, 
again with a fresh secret key. The new index is then used to replace the two indexes of size $t$.''

To provide real-world practicality to the scheme, I chose to modify this to avoid unnecessarily creating many tiny indexes
which will trigger many rapid successions of cascading merges with existing indexes. My design introduced the
notion of an SE index \emph{order} $o=\lceil\log N\rceil$. Rather than storing a single index $\cal I$, the server will
maintain a collection of indexes. Let ${\cal O}$ be the set of orders of indexes currently stored on a server. Then
${\cal C}=\{{\cal I}_o\ \forall o\in{\cal O}\}$ is the collection of all SE indexes which may logically appear to the client
as ``the SE index.'' 

When asked to perform a search, the server will perform the same operation on all 
indexes in $\cal C$, returning the union of their results.

For this scheme to work, I further impose the restriction that no two indexes of the same order may exist at the same
time in a collection (i.e., all elements of $\cal O$ are unique). When new data are to be added to the SE index, a new
index is built for the new data, resulting in a new index ${\cal I}_o$. If no other index of order $o$ exists in $\cal C$,
then ${\cal I}_o$ is added to $\cal C$. Otherwise, the contents of the existing ${\cal I}_o$ are merged with the new data
to create a new index ${\cal I}_p$. Then ${\cal I}_p$ \emph{replaces} the old ${\cal I}_o$ in $\cal C$. It may or may not be
the case that $o=p$. (If, at this point, there is an existing $p$-order index in $\cal C$, then this process is repeated
until all the cascading merges have resulted in an index of an order not currently on the server.)

By using this exponential progression in index sizes, I seek to minimize the amount of index data rebuilt at any given time.
The larger-order indexes (which will have more data in them) are merged less frequently than the smaller, lower-order ones.

Implementing this feature required a compromise to be added to the original scheme proposed by Demertzis and Papamantou---I 
added an encrypted list $\Delta$ of all indexed search terms. Only the client can decrypt this list, and it is never referenced
during normal operations. It is only accessed and decrypted by the client during merge operations. Doing this is necessary
because the SE index is set up to be useful only if a client already knows what search terms they're looking for, so there was
no previous reason to store $\Delta$ inside the index. Thus, no means were provided to reconstruct the original set $\Delta$ 
of search terms that was used. Without that information, the rest of the index cannot be decoded back into the original set 
of tuples.

My updated index-building process (replacing Algorithm~\ref{alg:setup}) is summarized in Algorithm~\ref{alg:merge}.

\begin{algorithm*}
 \begin{center}
  \begin{algorithmic}
   \Procedure{IndexGen}{$k,\cal D$}\Comment{Modifies and extends original \Call{Setup}{$k,\cal D$}}
    \State Let $\Delta$ be the list of all ``interesting'' keywords in document set $\cal D$.
    \State $N\gets\sum_{\forall w\in\Delta}|{\cal D}(w)|$
	\State $\ell\gets\lceil\log N\rceil$
    \State ${\cal L}\gets\{\ell,\ell-p,\ldots,\ell-(s-1)p\}$. 
    \State Let order $o=\ell$. We will now build a new order-$o$ index ${\cal I}$.
    \If{$L>1$} 
     \State ${\cal L}\gets{\cal L}\cup\{0\}$
    \EndIf
    \If{An order-$o$ index already exists on the server}
     \State Retrieve and decrypt $\Delta_o$ from server from existing order-$o$ index ${\cal I}_o$
     \State Retrieve and decrypt all storage buckets holding actual data from existing index ${\cal I}_o$ to ${\cal D}_o$
     \State ${\cal D}'\gets{\cal D}\cup{\cal D}_o$
	 \State Delete old index ${\cal I}_o$
     \State \Return\Call{IndexGen}{$k,{\cal D}'$}
	\EndIf
	\State $p\gets\lceil\ell/s\rceil$
    \State $\forall i\in\cal L$ organize storage level $A_i$, divided into buckets $A_i[x]$.
    \For{each keyword $w\in \Delta$ in random order}
    	\State Find adjacent $i,j\in{\cal L}:L2^j<|{\cal D}(w)|\le L2^i$.
    	\State Split ${\cal D}(w)$ into a set of chunks $C_w$. Set $c=0$.
    	\For{each chunk $v\in C_w$}
    		\State $c\gets c+1$.
    		\State Let $A$ be buckets in $A_i$ able to hold chunk $v$.
    		\State Pick one bucket $A_i[x]$ from $A$ randomly; store $v$ in it.
    		\State Add $H(F(k_1,w)\concat c)\Rightarrow [i\concat x]\oplus H(F(k_2,w)\concat c)$ to hash table.
    	\EndFor
   \EndFor
   \State Encrypt $\Delta$ and store in hash table in blocks of 100 words.
   \State Permute and encrypt entries in ${\cal L}$; fill HT with random data.
   \State Upload $\cal I$ to server.
   \EndProcedure
  \end{algorithmic}
  \caption{Create or Add to a Dynamic Index}\label{alg:merge}
 \end{center}
\end{algorithm*}

\section{Evaluation}
My evaluation of this implementation sought to verify the correct operation of the new dynamic indexing scheme,
and to measure its runtime efficiency for various runtime input sizes. In particular, I wanted to see the effects
of performing updates over different time intervals to see if there would be any benefit to updating the index
in real time, or daily, weekly, or less frequently.

The correctness was confirmed through a number of testcases which confirmed that the input data encoded into the
index was correctly retrieved again by search operations. The remainder of the evaluation is focused on performance
characteristics of the update function itself.

\subsection{Experimental Methodology}
To evaluate the efficiency of my implementation, I set up test cases based on real-world data samples representative
of highly dynamic document sets. I specifically chose email to approximate casual conversations which include both
message text and metadata headers. Indexing archives of chat rooms, text messages, and other similar communication
would be analogous.

My dataset was taken from the Enron email archive, \cite{klimt} which is a collection of
517,401 actual email messages belonging to 150 users.\footnote{I used a privacy-cleaned list which is slightly smaller
than the original release of the dataset.} Since this is a collection of actual communication between users, it provides
a valuable test case to simulate how my SE implementation would fare in a real application. As noted by Klimt
and Yang, ``It can be used both as a large sample of real life email users and as a standard
corpus for comparison of results using different methods.'' \cite{klimt2}

For the purposes of the evaluation, each email message is stored in an individual disk file (in Maildir format).
The document ID ($\textsf{id}(D)$) for each is the system's inode number for the message file. This provides a
guaranteed unique ID for each file without the overhead of assigning custom IDs.

One experimental run of my dynamic SE implementation looked at the case of maintaining a comprehensive index of all
messages. To simulate a realistic flow of new messages coming into the repository, I batched up the messages according
o the original \z{Date:} header lines.

\begin{figure*}
 \includegraphics[width=\textwidth]{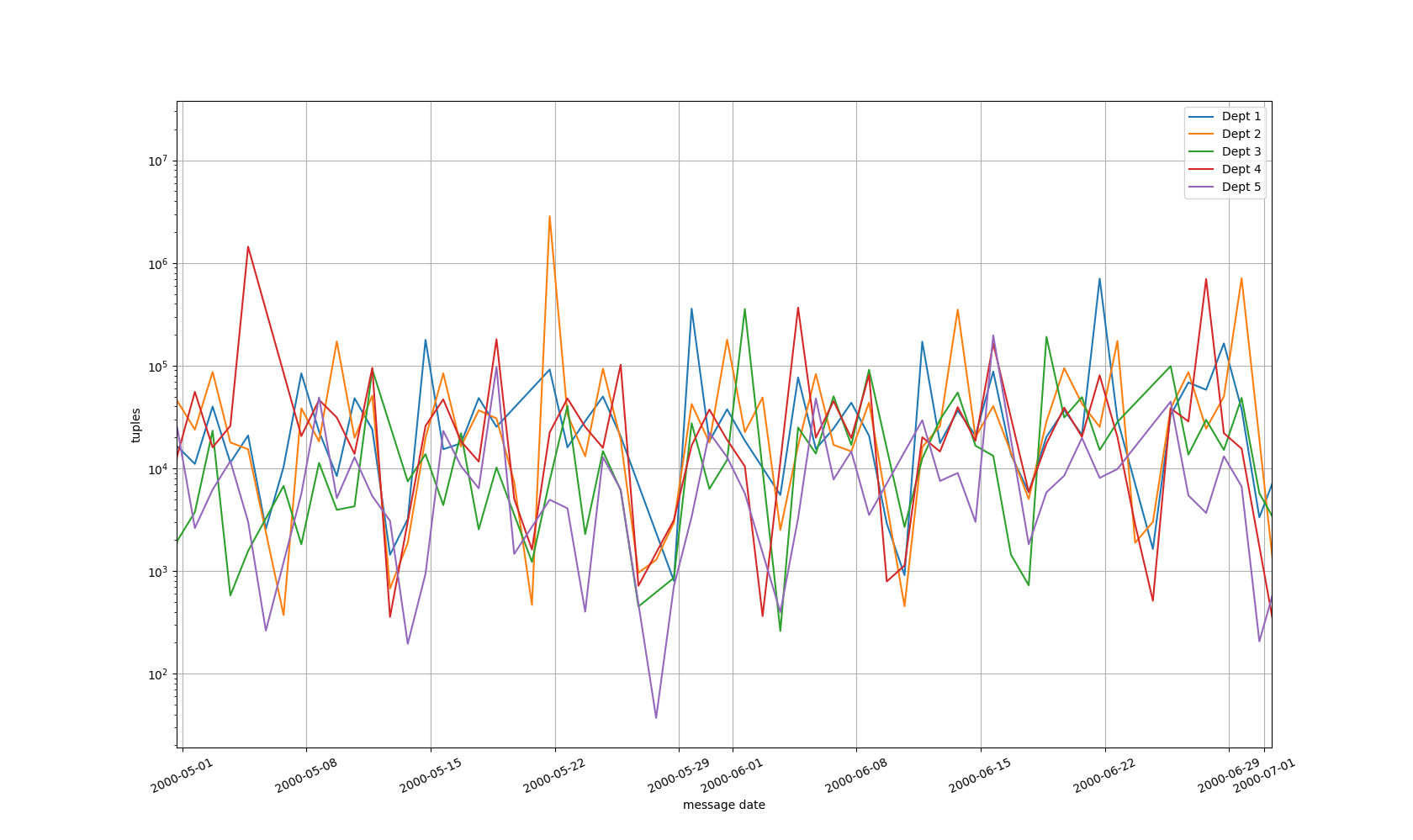}
 \caption{Incoming Keyword Tuples For All Departments\label{fig:all_depts_closeup}}
\end{figure*}

I ran three separate experiments, using batch sizes of 1, 7, and 30 days, to measure the performance of
the implementation if it were to be re-indexing on a daily, weekly, or monthly basis respectively.
Each of these operated on five subsets of the Enron data corpus, organized as shown in Table~\ref{tbl:depts}
by the recipient's initials. This is meant to simulate an arbitrary partitioning of a workforce into ``departments''
which may have slightly different input patterns.  Figure~\ref{fig:all_depts_closeup} shows the day-by-day intake
of $(w,\textsf{id}(D))$ tuples for each of the departments. We see from this that although there are differences in
activity day-to-day, the overall pattern of activity was similar, giving us five sample sets to compare and contrast.
I observed consistent behavior among all five departments as I examined the server resource usage as each of their
document indexes expanded over time.

\begin{table*}
 \begin{center}
  \begin{tabular}{cr@{--}lrrr}\toprule
   \bfseries Department & \multicolumn{2}{c}{\bfseries Initial} & \bfseries People & \bfseries Messages & \bfseries Data (kB) \\\midrule
   1 & A&F & 33 & 119,477 & 667,632 \\
   2 & G&K & 29 & 143,652 & 760,388 \\
   3 & L&P & 32 & 103,354 & 504,668 \\
   4 & Q&S & 38 & 105,930 & 526,084 \\
   5 & T&Z & 18 &  44,988 & 225,600 \\\bottomrule
  \end{tabular}
  \caption{Arrangement of Users into Departments\label{tbl:depts}}
 \end{center}
\end{table*}

With the input data thus broken into batches of various sizes, I ran each set of updates
while measuring the following performance characteristics after each update operation:
\begin{itemize}
 \item Size $N$ of each index ${\cal I}_o$ in terms of $N$
 \item Disk storage on the server for each ${\cal I}_o$
 \item Full set of words $\Delta$ added in that update
 \item Contents of all storage locations in $A_i$
 \item Histogram of distribution of each keyword $w$ within the input document set $\cal D$
 \item Wall-clock time in seconds taken to build the new index (including any needed merges with existing ones)
 \item Number of network transactions required to perform the update operation
 \item Number of bytes exchanged between client and server during the update operation
 \item Number of input documents added during that update
 \item If merging, how many words and tuples from previous indexes were downloaded to be merged into the new index
 \item Which orders of indexes were merged at each update
\end{itemize}

\subsection{Results}
\begin{figure*}
 \includegraphics[width=\textwidth]{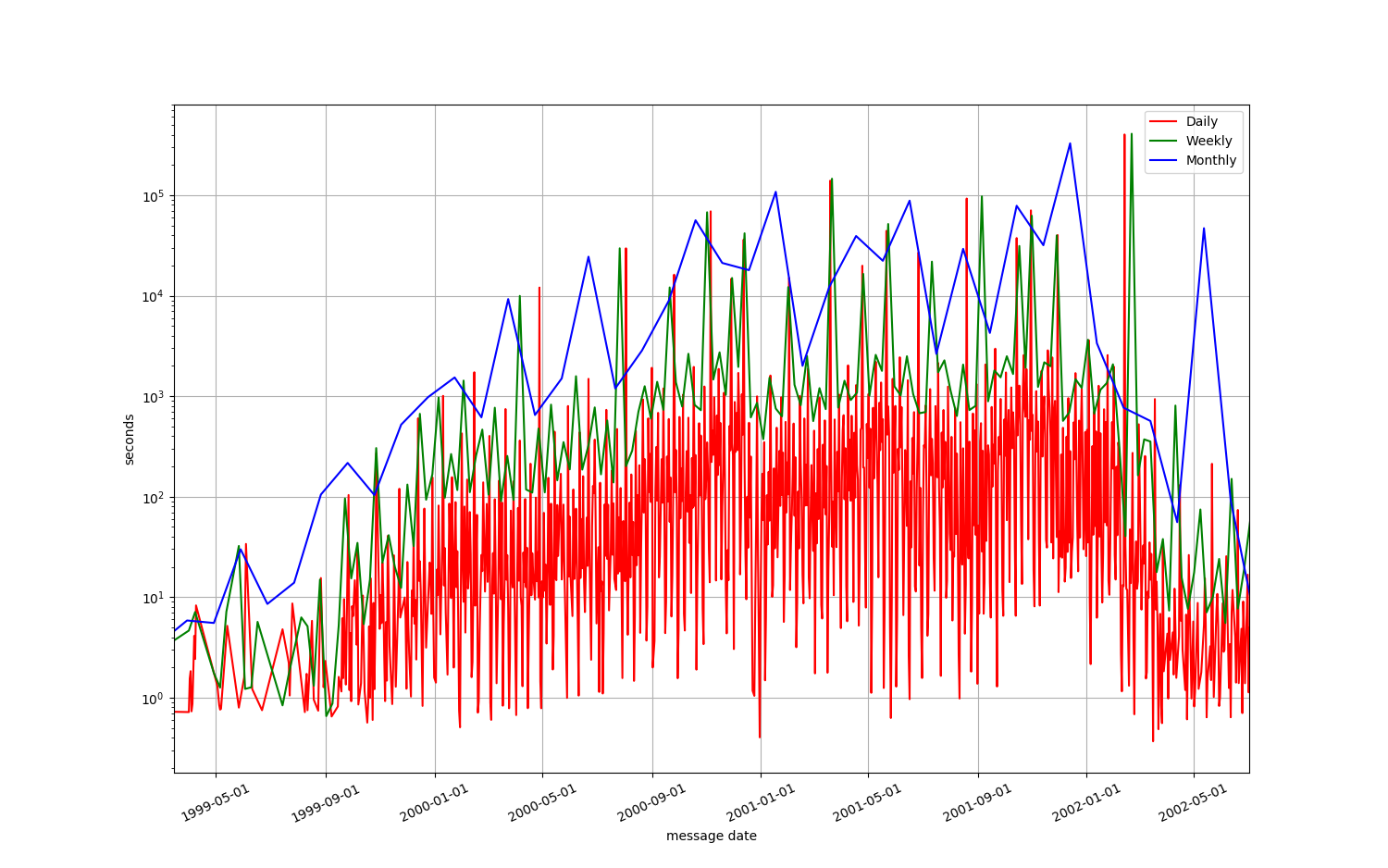}
 \caption{Processing Time (Dept.~1) by Batch Size\label{fig:DWM}}
\end{figure*}

\begin{figure*}
 \includegraphics[width=\textwidth]{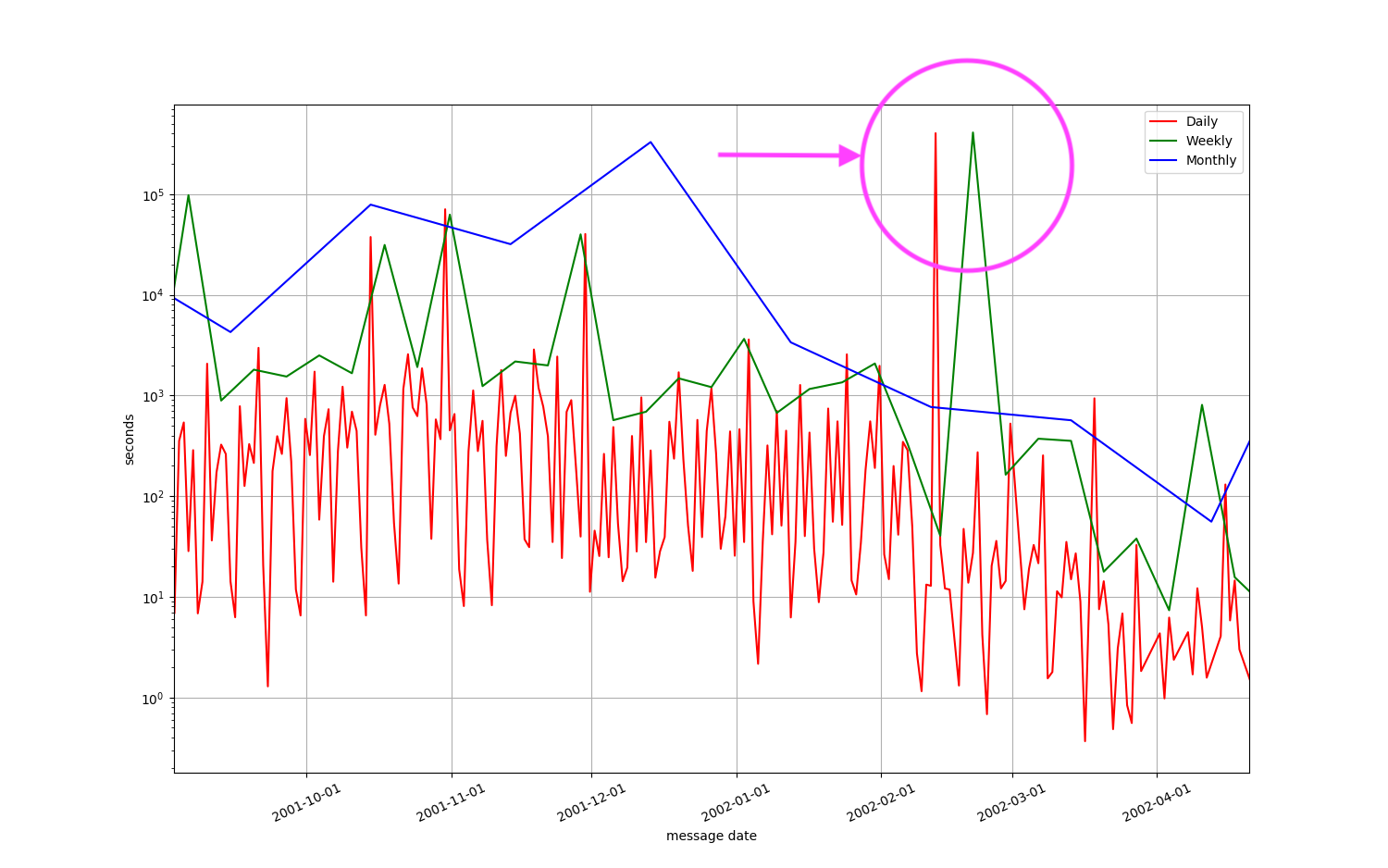}
 \caption{Processing Time (Dept.~1) by Batch Size, Detail View\label{fig:DWM_closeup}}
\end{figure*}

Using Department~1 as a representative example of the results obtained, we see in Figure~\ref{fig:DWM} how the
processing time varied as a function of the frequency of updates. At first glance, it is apparent that we can
get an overall savings in processing work by doing updates less frequently. For example, in the close-up view
in Figure~\ref{fig:DWM_closeup} we see that over the same period of time the daily and weekly updates saw
spikes of activity as they had to merge multiple indexes but the monthly updates (blue line) did not, since it
had the advantage of performing the update operation with more data at a time locally in a single step rather than
making more incremental updates which needed to be downloaded again to be merged later.

This prompted me to seek a predictor for these merge events, as a potential avenue to optimize the merge by
anticipating when merges would be necessary. While there are some straightforward (if na\"\i ve) indicators
we could use, such as the number of incoming documents $|{\cal D}|$ or the number of input words $|\Delta|$
or even the number $N$ of incoming tuples, none of these is a completely accurate predictor of either the
time a merge will happen, or of the magnitude of each merge event. 

\begin{figure*}
 \includegraphics[width=\textwidth]{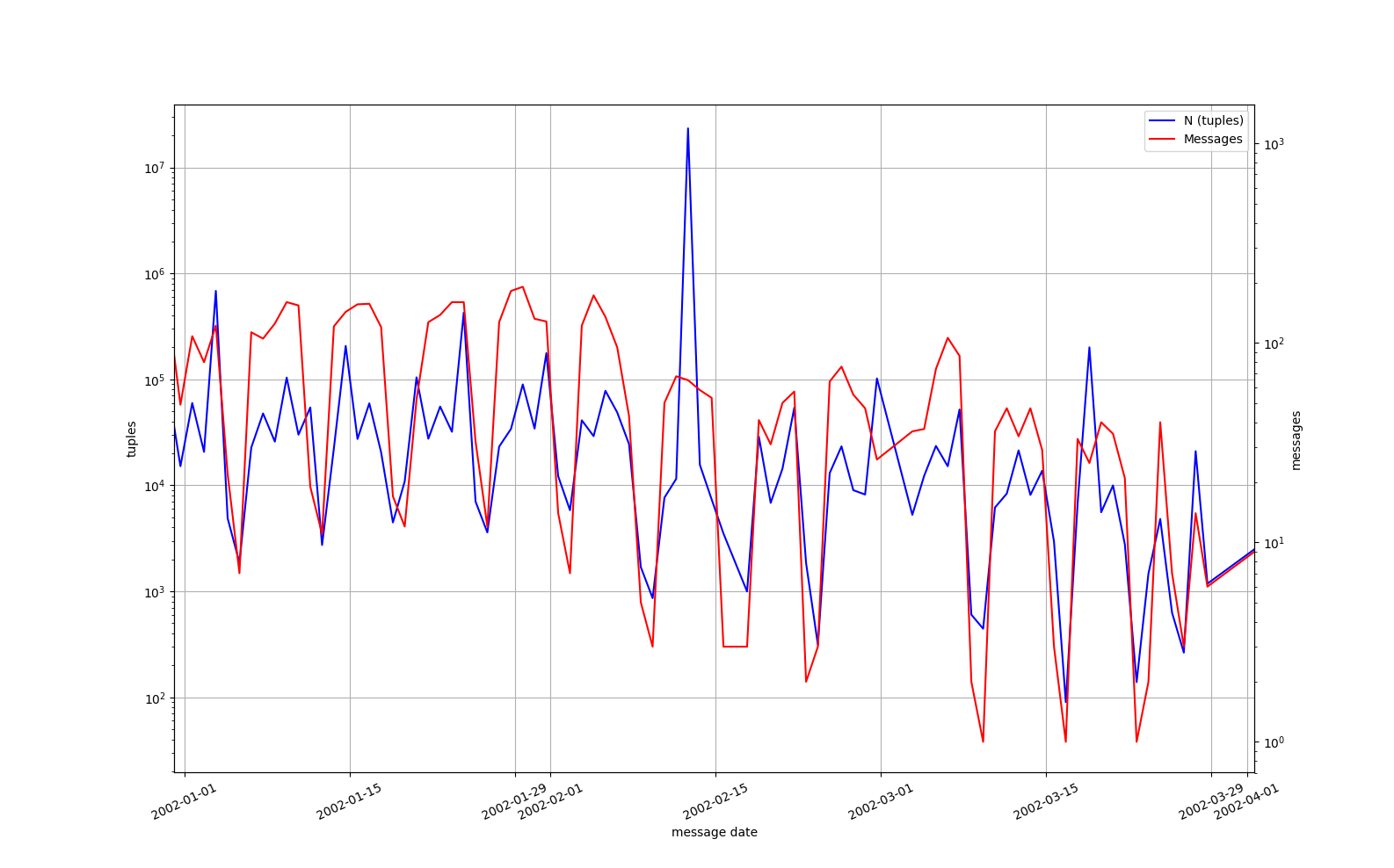}
 \caption{Incoming Tuple Count vs.\@ Message Count (Dept.~1)\label{fig:Dept1_n_msg_closeup}}
\end{figure*}

The reason for this is that the conditions which trigger a merge event depend on the exact distribution of
keywords ($w$) in each existing index ${\cal I}_o\in{\cal C}$ as well as how the specific incoming data will
alter that distribution. As we see in the data sample in Figure~\ref{fig:Dept1_n_msg_closeup}, a given input
batch may have a few messages with a high diversity of input words (driving $N$ significantly higher than $|{\cal D}|$)
or vice versa.

If we want a loose correlation to track with the performance over time of the SE updates, $N$ would still seem
the best reasonable value that is easily at hand, which may be useful for long-range statistical analysis
including the prediction of incoming message volume to the system overall, from which merge probability may be
inferred.

\subsection{Issues}
I discovered a pathological condition as an index hit 
certain sizes requiring large-scale cascading re-indexing operations. Most updates were completing in seconds,
but these exceptional cases were taking many hours.
In in some cases they took days to complete. For example, among the daily updates for Dept.~1 there were 3 out of the 
total 995 update batches (0.3\% of the total) which took more than one day to complete. These took 
1 day, 1 hour;
1 day, 14 hours; and
4 days, 15 hours.

Further investigation led to an initial diagnosis that this was likely caused by a combination of resource
scarcity on the server and inefficiencies in the Python implementation which still need to be optimized further.
For example, at some points an index collection included an order-24 index. This would contain a small number of
buckets of size 33,554,432 words. For 64-bit words (as my implementation uses to store the $i\concat x$ encoded 
values), that's approximately a quarter-gigabyte string value to be assembled, copied, transmitted, received, buffered,
copied, and decoded. In a runtime system such as Python's, that is not necessarily going to be handled as efficiently
as it might be.

\subsection{Summary of Results}
Overall, the results indicate that even this experimental Python implementation performed well enough to be
of practical use. Updates finished in time for the index to be used for a reasonable time before the next update
batch needed to start. The rare exceptions to this were due to the pathological case described in the previous
section. Table~\ref{tbl:summary} summarizes the runtime results of each of the experiments.

\begin{table*}
 \begin{center}
  \begin{tabular}{lrrrrrrrrrr}\toprule
   & \multicolumn{2}{c}{\bfseries Dept.~1} 
   & \multicolumn{2}{c}{\bfseries Dept.~2} 
   & \multicolumn{2}{c}{\bfseries Dept.~3} 
   & \multicolumn{2}{c}{\bfseries Dept.~4} 
   & \multicolumn{2}{c}{\bfseries Dept.~5}\\
   \bfseries Sample &
   \bfseries Avg & \bfseries SD &
   \bfseries Avg & \bfseries SD &
   \bfseries Avg & \bfseries SD &
   \bfseries Avg & \bfseries SD &
   \bfseries Avg & \bfseries SD \\\midrule
   Daily &  21.14 & 240.75 &  4.75 &  47.87 &  11.07 & 111.57 &  12.49 & 119.33 &  0.99 &  11.20 \\
   Weekly& 110.92 & 576.32 & 25.28 & 117.14 &  53.82 & 242.18 &  72.53 & 301.46 & ---   & ---   \\
   Monthly&322.31 & 855.13 & ---   & ---    & 197.31 & 504.38 & 254.61 & 583.68 & 96.26 & 341.74 \\
   \bottomrule
  \end{tabular}
  \caption{Summary of Experimental Results (batch wall-clock time in minutes)\label{tbl:summary}}
 \end{center}
\end{table*}

\section{Future Work}
Given the success of the experimental Python implementation, it makes sense to continue optimizing this design
by coding it in a more efficient runtime system based on a language such as C or Go, as well as to continue looking
for ways to optimize the server protocols and possibly the storage system itself. Specifically, the cause of the
occasional very-long update operations should be investigated more.

This system also needs to be expanded to include the concept of deletion of messages from the index.

Finally, a formal evaluation of the cryptographic strength, including the likelihood and impact of potential
information leakage over time when using this design compared to other SE schemes.

I did not examine the effects of changing the locality and storage variables $L$ and $s$ since that was already
thoroughly treated by Demertzis and Papamanthou \cite{demertzis} in their proposal for this SE architecture
initially. However, it would be interesting to come back to that once my dynamic changes to their design
have matured and evolved into a fully workable model with deletion support, to see if then the effect of adjusting
those variables is different from what was found previously.

\section{Conclusions}
I have expanded on the work of previous SE researchers to implement an experimental yet functional dynamic
SE indexing service. With that in place, I have analyzed the runtime performance with the update batch size
as the independent variable for my experiments. I concluded from those experiments that the scheme as described
in this paper is practical for real-world applications. Further, I identified how
more efficient
updates (requiring less overall work) are achieved by delaying updates for longer periods of time (e.g.,
weekly or monthly rather than hourly or daily). However, this comes at the cost of not having that new data
available for users of the index. It is necessary for a server administrator to determine what update frequency
serves the needs of their users best.

This work contributes to the future implementation of real-world encrypted document indexing systems which
can be employed to organize repositories of chat logs, emails, discussion forums, or other dynamic
document collections while protecting the confidentiality of the content being served.

\section{Acknowledgements}
I wish to express gratitude to the guidance provided by Dr.\@ Charles V. Wright, my Ph.D. advisor, who supervised this
research and provided valuable feedback along the way. Also to Dr.\@ David Maier, for his advice and instruction on writing
styles for the earliest drafts of this paper.

\bibliographystyle{unsrt}
\bibliography{\jobname}

\begin{thebibliography}{10}

\bibitem{curtmola}
Reza Curtmola, Juan Garay, Seny Kamara, and Rafail Ostrovsky.
\newblock Searchable symmetric encryption: improved definitions and efficient
  constructions.
\newblock In {\em Proceedings of the 13th ACM conference on Computer and
  communications security}, pages 79--88, 2006.

\bibitem{demertzis}
I.~Demertzis and C.~Papamanthou.
\newblock Fast {S}earchable {E}ncryption with {T}unable {L}ocality.
\newblock In {\em Proc.\@ 2017 ACM Int.\@ Conf.\@ Management of Data}, pages
  1053--1067, 2017.

\bibitem{song}
Dawn~Xiaoding Song, David Wagner, and Adrian Perrig.
\newblock Practical techniques for searches on encrypted data.
\newblock In {\em Proceeding 2000 IEEE symposium on security and privacy. S\&P
  2000}, pages 44--55. IEEE, 2000.

\bibitem{pinkas}
Benny Pinkas and Tzachy Reinman.
\newblock Oblivious ram revisited.
\newblock In {\em Advances in Cryptology--CRYPTO 2010: 30th Annual Cryptology
  Conference, Santa Barbara, CA, USA, August 15-19, 2010. Proceedings 30},
  pages 502--519. Springer, 2010.

\bibitem{goldreich}
Oded Goldreich and Rafail Ostrovsky.
\newblock Software protection and simulation on oblivious rams.
\newblock {\em Journal of the ACM (JACM)}, 43(3):431--473, 1996.

\bibitem{cash}
David Cash, Joseph Jaeger, Stanislaw Jarecki, Charanjit Jutla, Hugo Krawczyk,
  Marcel-C{\u{a}}t{\u{a}}lin Ro{\c{s}}u, and Michael Steiner.
\newblock Dynamic searchable encryption in very-large databases: Data
  structures and implementation.
\newblock {\em Cryptology ePrint Archive}, 2014.

\bibitem{cash2}
David Cash and Stefano Tessaro.
\newblock The locality of searchable symmetric encryption.
\newblock In {\em Advances in Cryptology--EUROCRYPT 2014: 33rd Annual
  International Conference on the Theory and Applications of Cryptographic
  Techniques, Copenhagen, Denmark, May 11-15, 2014. Proceedings 33}, pages
  351--368. Springer, 2014.

\bibitem{asharov}
Gilad Asharov, Moni Naor, Gil Segev, and Ido Shahaf.
\newblock Searchable symmetric encryption: optimal locality in linear space via
  two-dimensional balanced allocations.
\newblock In {\em Proceedings of the forty-eighth annual ACM symposium on
  Theory of Computing}, pages 1101--1114, 2016.

\bibitem{naveed}
Muhammad Naveed, Seny Kamara, and Charles~V Wright.
\newblock Inference attacks on property-preserving encrypted databases.
\newblock In {\em Proceedings of the 22nd ACM SIGSAC Conference on Computer and
  Communications Security}, pages 644--655, 2015.

\bibitem{zhang}
Yupeng Zhang, Jonathan Katz, and Charalampos Papamanthou.
\newblock All your queries are belong to us: the power of
  $\{$File-Injection$\}$ attacks on searchable encryption.
\newblock In {\em 25th USENIX Security Symposium (USENIX Security 16)}, pages
  707--720, 2016.

\bibitem{ipfs}
Juan Benet.
\newblock Ipfs-content addressed, versioned, p2p file system.
\newblock {\em arXiv preprint arXiv:1407.3561}, 2014.

\bibitem{klimt}
Bryan Klimt and Yiming Yang.
\newblock Introducing the enron corpus.
\newblock In {\em CEAS}, volume~45, pages 92--96, 2004.

\bibitem{klimt2}
Bryan Klimt and Yiming Yang.
\newblock The enron corpus: A new dataset for email classification research.
\newblock In {\em European conference on machine learning}, pages 217--226.
  Springer, 2004.

\end{thebibliography}

\end{document}